# Evaluation of Linear Implicit Quantized State System method for analyzing mission performance of power systems

**Navid Gholizadeh[1], Joseph M Hood[1], and Roger A Dougal[1]**

## Abstract

The Linear Implicit Quantized State System (LIQSS) method has been evaluated for suitability in modeling and simulation of long duration mission profiles of Naval power systems which are typically characterized by stiff, non-linear, differential algebraic equations. A reference electromechanical system consists of an electric machine connected to a torque source on the shaft end and to an electric grid at its electrical terminals. The system is highly non-linear and has widely varying rate constants; at a typical steady state operating point, the electrical and electromechanical time constants differ by three orders of magnitude—being 3.2 ms and 2.7 s respectively. Two important characteristics of the simulation— accuracy and computational intensity—both depend on quantization size of the system state variables. At a quantization size of about 1% of a variable's maximum value, results from the LIQSS1 method differed by less than 1% from results computed by well-known continuous-system state-space methods. The computational efficiency of the LIQSS1 method increased logarithmically with increasing quantization size, without significant loss of accuracy, up to some particular quantization size, beyond which the error increased rapidly. For the particular system under study, a "sweet spot" was found at a particular quantum size that yielded both high computational efficiency and good accuracy.

## Definitions

**Reference system**: An electromechanical machine driven by a torque source and coupled to an ideal electrical power bus (i.e. zero-impedance voltage source)

**Reference model**: The set of differential-algebraic equations that describe the time-domain behavior of the *Reference System* as formulated in a reference frame rotating synchronously with the electric grid according to the Park transformation.

**Reference solution**: The solution of the *Reference Model* obtained by applying the Euler method

**State update intensity**: The number of updates to any quantized state variable per unit of time in the simulation frame

**Pointwise error (PE)**: At any time instant, the difference between the value $y_{ij}$ of a state variable that is computed by the LIQSS method resampled at the time step of the reference simulation and the value $q_{ij}$ computed by a known-good reference simulation, i.e., $PE = y_{ij} - q_{ij}$

**Time average normalized error (TANE)**: The square root of the mean value over time of the squared PE resampled at the time step of the reference simulation and normalized by the dynamic range of $y_i$ during the period of interest

---

[1]University of South Carolina, USA

**Corresponding author:**
Roger A Dougal, University of South Carolina, 301 Main Street, Columbia, SC 29208, USA.
Email: Dougal@cec.sc.edu

$$\textit{Time Average Normalized Error}_i = \frac{\sqrt{\sum_j^N \frac{(PE)^2}{N}}}{max(\mathbf{y_i}) - min(\mathbf{y_i})} \quad (1)$$

$TANE_i$: TANE in the $i$th state variable

$N$: The number of uniform (resampled) time steps from the beginning to the end of the simulation

**Maximum error:** The largest of the TANE among the Average Errors of all system states

**Quantum size or quantization size:** The range of a continuous variable that is represented by a single discrete value after quantization. A constant that defines the set of discrete values the output of a state variable quantizer may assume. In other words, the output value of a state variable is $\Delta Q \cdot k$, where $\Delta Q$ is the quantization size and $k$ is an integer ($k \in \mathbb{Z}$).

**QSS atom**: A computational unit (or programming object) that stores and updates a single state variable's internal continuous value and external output using QSS method.

**QSS**: Quantized State System, a system comprising one or more connected QSS atoms.

**LIQSS**: A solution method that uses linear implicit integration to estimate the time to reach the next higher or lower quantized state.

**LIQSS solution**: The solution of a system model obtained by applying the LIQSS method.

## Introduction

The purpose of the research reported here was to assess the performance of the Linear Implicit Quantized State System (LIQSS) method[1] when applied to the studies of the dynamic performance of power and energy systems (PES). Naval PES typically operate for long durations in quasi-steady states, but these steady states are occasionally interrupted by transient events having relatively fast dynamics. This study takes as reference a synchronous electromechanical machine (a motor or generator), as one important representative of the general class of non-linear components that comprise a typical Naval PES.

This work builds on other earlier attempts at simulating electrical circuits using various QSS-based methods. Literature[2] describes the application of QSS1 to solve a tenth-order linear system and QSS2 to solve a second-order non-linear system, but does not reach to higher orders of non-linear systems. The QSS2 solution, when tuned to require a comparable number of steps as Heun's method (classic second-order fixed-step method), produced smaller maximum error (relative to the true trajectory) than did Heun's method. In Hood and Dougal,[3] a dense, linear network having a very high stiffness ratio ($10^9$) was simulated using a combination of LIQSS and latency methods.[4] The so-called quantized devs with latency (QDL) method successfully solved the system response even though conventional methods could not (or would have required impractically long times to do so due to the need for small time steps to ensure stability yet the need to cover a long duration to reveal the slow part of the system response). That work did not address the implications of non-linear system elements. In Di Pietro et al.,[1] a four-stage interleaved converter was simulated using a modified version of LIQSS2 (mLIQSS2). In all of these previous works, the feasibility and performance of QSS-based methods were investigated only for linear systems or for very small non-linear systems.

The key contribution of this paper is in describing the performance of LIQSS1 when simulating a higher order—even if not yet large—non-linear power system. This investigation is a prelude to future work that will study the use of latency methods as a means of enforcing algebraic constraints in non-linear systems, but we believe the results pertaining to the LIQSS1 performance are interesting in and of themselves.

This study is motivated by the fact that synchronous machines in naval power systems often operate for long durations in relatively steady conditions, but these steady conditions are occasionally punctuated by abrupt changes of conditions that must be accurately characterized. QSS methods appear to offer high computational efficiency for simulation of such systems, and we are interested in variations of these methods that will improve the performance when applied to the studies of non-linear PES. Non-linearity, high stiffness, and a necessity for algebraic constraints to enforce circuit conservation laws are the important characteristics of these systems, and they are exhibited in the study system reported here.

## Quantized state system methods

The Quantized Discrete Event Specification (QDEVS)[5] provides a formal DEVS specification for a discrete event description of quantized systems. The Quantized State System (QSS) methods are a series of integration methods based on the QDEVS specification and described in Zeigler et al.[5] These QSS methods provide a QDEVS-compliant way to simulate continuous systems.

The QSS approach begins with the assumption that a generic continuous state equation system (SES)

$$\dot{x} = f(\mathbf{x}(t), u(t)) \quad (2)$$

can be approximated by a QSS in the form

$$\dot{x} = f(\mathbf{q}(t), u(t)) \quad (3)$$

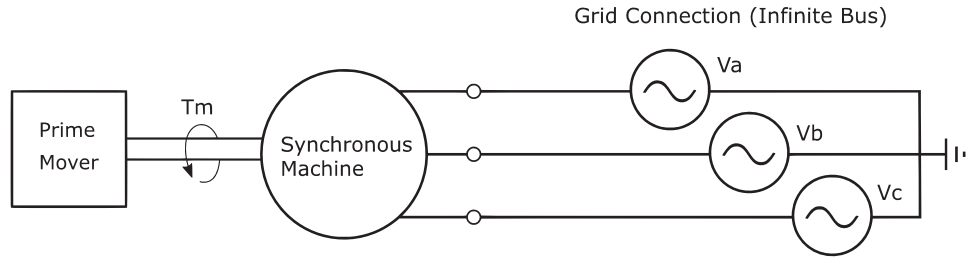

**Figure 1.** Synchronous generator connected to an infinite bus.

where **q** is the quantized state vector that follows piecewise constant trajectories and is related to the state vector **x** by the quantum size $\Delta Q$.

Literature[6,7] defines the structure and implementation of atomic DEVS models and general-purpose simulators for QSS systems. The QSS approach guarantees a bounded error,[8] so analytically stable systems cannot become numerically unstable when being simulated by a fully coupled QSS algorithm.[7]

Several variations of QSS offer different features. The simplest formulation, QSS1, developed in Zeigler et al.[5] and Kofman and Junco,[8] relies on explicit integration and uses first-order estimates of state derivatives to predict the time at which the continuous state $x_i(t)$ will increase or decrease by amount $\Delta Q$ (quantization step size) from the current quantized value $q_i(t)$ to the next higher or lower quantized value. Although QSS1 has some advantages, like being easy to implement, its disadvantage is that it uses a first-order approximation of the state trajectory to calculate the time to the next event; to get accurate results, $\Delta Q$ has to be quite small, which produces a large number of steps. QSS2[2] and QSS3[9] use more accurate second- and third-order approximations, respectively, for the state trajectory, however, the computational cost grows with the square root and cubic root of the desired accuracy.

Because we are interested in simulating realistic, stiff power systems that include both fast electrical dynamics and slow mechanical dynamics, we require a QSS method that can handle stiff systems. This requirement eliminates QSS1, QSS2, and QSS3 because these methods create fictitious high-frequency oscillations[1] which in turn generate large numbers of steps that are costly in computational cost and memory size, even when a system is nominally in steady state. We chose instead to use the LIQSS methods which were specifically developed to address the concurrent existence of slow and fast dynamics inherent in stiff systems.[1]

LIQSS methods combine classic implicit integration techniques into the QSS methods. Similar to the way that several variations of QSS methods were developed, so also were variations of LIQSS, such as LIQSS1, LIQSS2, and LIQSS3. These perform first-, second-, and third-order approximations, respectively.[1] The LIQSS2,[1,10] MLIQSS2,[1] and LIQSS3 methods all offer improvements and performance and stability over the original LIQSS.

Despite the benefits of LIQSS2 or 3, the simplicity of LIQSS1[1,10] compelled us to use it in this study where our focus is on how computational intensity scales with quantization size for non-linear systems of order higher than 2, (second-order systems were already reported in Kofman[2]). Furthermore, in future work, we intend to report the performance of the combination of LIQSS1 with latency insertion method (LIM),[4] or QDL,[3] when solving stiff non-linear systems. Not only will LIQSS1 make it easier to implement the necessary models but we also anticipate that using the first-order method will make it easier to distinguish latency effects from integration effects. If latency methods usefully improve simulator performance for first-order methods, then extensions can later be made to higher order variations of LIQSS, perhaps with additional gains in performance and stability.

## Reference electric power system

The reference system, shown in Figure 1, has three major components—a prime mover, a synchronous machine (which can act as either a generator or a motor depending on the direction of power flow), and an AC power grid. Models of the prime mover and the grid are simplified—the torque source is an ideal time-dependent source with zero inertia, and the power system is represented as an ideal three-phase sinusoidal AC voltage source, with zero impedance and constant frequency—but these simplifications do not limit the general applicability of our results because the machine model still entails the solution of non-ideal network equations. Our particular reference system was chosen because it is of widespread interest, and because it also demonstrates suitability of the QSS method to efficiently solve systems that are characterized by coupled fast and slow dynamics.

## Reference model

The model of the electric machine is formulated in the synchronous reference frame to eliminate the periodic sinusoidal variations of all voltages and currents. This use of the standard Park transformation[11] maximizes the value of the QSS method for the analysis of AC systems. Although QSS can be used to simulate systems having arbitrary waveforms, the sinusoidal voltage and current oscillations inherent in an AC power system, if not factored out by a method such as Park transformation, would require rapid state updates that would obviate any benefits offered by the QSS method.

A common model of the synchronous generator, when formulated in the synchronous reference frame of the Park transformation, is represented by the set of equivalent circuits shown in Figure 2. For the system of our analysis, the basis frequency is 50 Hz. The seventh-order set of nonlinear equations that describe the dynamics of the transformed circuit is described in Equations (3)–(9), and the algebraic constraints that apply to the network solution are defined by Equations (10) and (11). Figure 2 and the following equations follow.[12]

The equations that describe these circuits are as follows

$$\frac{d}{dt}\psi_d = V_d(t) - R_s i_d + \omega_r \psi_q \tag{4}$$

$$\frac{d}{dt}\psi_q = V_q(t) - R_s i_q - \omega_r \psi_d \tag{5}$$

$$\frac{d}{dt}\psi_F = e_{fd} - i_F R_F \tag{6}$$

$$\frac{d}{dt}\psi_D = -i_D R_D \tag{7}$$

$$\frac{d}{dt}\psi_Q = -i_Q R_Q \tag{8}$$

$$\frac{d}{dt}\omega_r = \frac{n}{J}\left(i_q\psi_d - i_d\psi_q - T_m\right) \tag{9}$$

$$\frac{d}{dt}\theta = \omega_r - \omega_b \tag{10}$$

$$\begin{bmatrix} i_{dr} \\ i_F \\ i_D \end{bmatrix} = \begin{bmatrix} L_{md}+L_L & L_{md} & L_{md} \\ L_{md} & L_F+L_{md} & L_{md} \\ L_{md} & L_F+L_{md} & L_F+L_{md} \end{bmatrix}^{-1} \cdot \begin{bmatrix} \psi_{dr} \\ \psi_F \\ \psi_D \end{bmatrix} \tag{11}$$

$$\begin{bmatrix} i_{qr} \\ i_Q \end{bmatrix} = \begin{bmatrix} L_{mq}+L_L & L_{mq} \\ L_{mq} & L_{mq} \end{bmatrix}^{-1} \cdot \begin{bmatrix} \psi_q \\ \psi_Q \end{bmatrix} \tag{12}$$

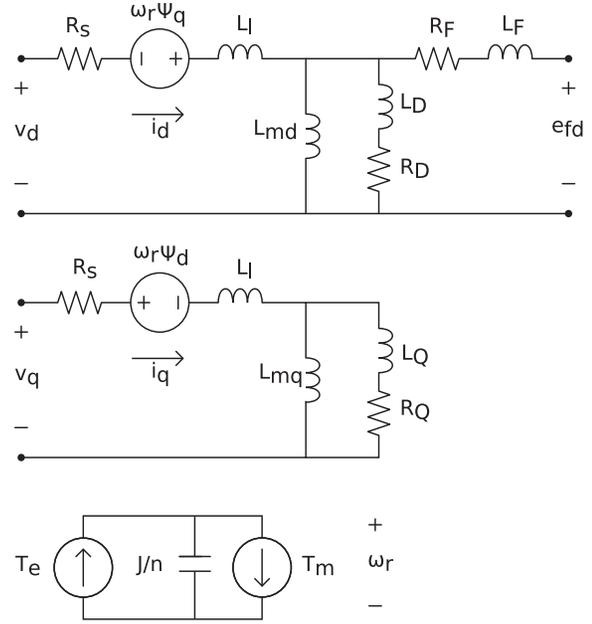

**Figure 2.** Direct, quadrature, and mechanical equivalent circuits of the synchronous machine.

where the terms in these equations are defined as follows:

$\psi_d, \psi_q$: direct and quadrature stator fluxes in q, d equivalent circuits, respectively

$\psi_D, \psi_Q$: direct and quadrature damper fluxes

$V_d(t), V_q(t)$: direct and quadrature components of the stator terminal voltages, respectively

$R_s$: series resistance in both d, q stator equivalent circuits

$\omega_r, \omega_b$: rotor speed in rad/s and base frequency $(2\pi f)$, respectively with f = 50 Hz

$i_d, i_q$: direct and quadrature components of currents in the d, q stator equivalent circuits

$\psi_F, i_F$: stator field flux and field current, respectively

$e_{fd}$: direct component of the field voltage

$R_F$: field resistance

$i_D, i_Q$: direct and quadrature internal equivalent currents

$R_D, R_Q$: d and q components of equivalent circuit resistance, respectively

$i_{dr}, i_{qr}$: d- and q-axis rotor currents

$\theta$: rotor angle relative to synchronous reference frame

Two solutions to the *reference model*, when operated through the *reference scenario*, were developed—a reference solution and a solution by the LIQSS method. The reference solution was obtained by applying Euler forward integration to the state-space equations, using MATLAB as the tool. A fixed time step of $10^{-4}$ s was chosen for the explicit Euler solution so that that all eigenvalues would be stable over the entire operating range of the system.

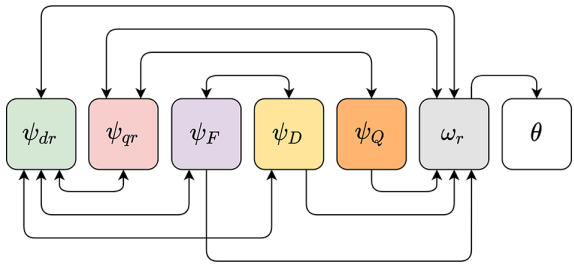

**Figure 3.** Atoms with lines showing which atoms trigger the updates in the other ones.

The second solution was obtained using the LIQSS method as described next. The code and the model parameters for both solutions are publicly available in a GitHub repository at the URL https://github.com/UofSC-QDEVS/LIQSS_On_NonlinearSys.

## Reference scenario

The reference system was exercised through the following operating scenario to produce all of the data reported here. The scenario starts with the synchronous generator spinning in steady state at 3000 r/min in synchronism with the sinusoidal grid voltage of 20 kV$_{rms}$ line-to-line. The machine produces an open circuit voltage equal in magnitude and phase to that of the grid. Since the phase angle ($\delta$) between the vector of power source voltage and the vector of stator voltage is zero, no current flows between the machine and the electric grid, so neither real (or "active") nor imaginary (or "reactive") power flows between the two, and zero torque is required to maintain the rotational speed. This steady state condition continues for the first 15 s. Beginning at $t = 15$ s, the torque applied by the prime mover to the shaft of the machine begins to ramp up. The torque ramp continues until $t = 20$ s, at which time the torque has reached 25% of the machine's rated torque. At 25% of rated torque, the phase of the stator voltage leads that of the grid voltage, and the machine drives 83 MW (real power, or active power) and 13 MVAR (imaginary power, or reactive power) into the grid.

## Implementation of the LIQSS1 model

The LIQSS1 model was formulated based on the specification in Di Pietro et al.[1] and Migoni and Kofman.[10] According to this method, any QSS atom computes the time at which it will reach a next state, and that is the time when the atom is next updated, unless an earlier update is required because an input has changed.

Figure 3 shows the seven QSS atoms used in modeling the synchronous machine. In this implementation, every input to any atom comes from an output that has been quantized. The lines with direction arrows indicate that the output of one atom is conveyed to the input of another atom. Since the system is tightly coupled, many components have bidirectional arrows. Iteration between the atoms represents a sort of relaxation process. Bidirectional arrows indicate that a state update of either atom requires an update of the other. As an example, following the string of just one arrow, any update to the output of $\psi_{dr}$ requires $\omega_r$ to update the time at which it expects to reach its next quantum level. If the update of $\psi_{dr}$ results in a change of the quantized state of $\omega_r$, then the $\theta$ atom must update the estimate of its time to the next quantum transition that will feed back to require a next update of $\psi_{dr}$. To start the solution, the "next update" time of each atom is initialized to infinity. Then, each atom computes its own next update time—the time at which it should arrive at its next higher or lower quantum state. The atom with the closest (smallest) update time is then updated. The occurrence of this update flags the atoms connected to it to again update their own time to the next quantum state, and the loop continues. A flowchart of the process is shown in Hood and Dougal.[3]

## LIQSS performance

The performance of the QSS method is shown in comparison to the performance of the reference method in a series of plots. Each plot shows the trajectory of a state variable and the number of updates of the computational atom for that state variable. All states in the system use the same quantum size ($\Delta Q$) of $10^{-4}$ Wb, except the machine rotor speed ($\omega_r$) for which we choose a quantum size 1/10 as large ($10^{-5}$ rad/s) since the dynamic range of the rotor speed is much less than that of the system fluxes. We have not established a mathematically rigorous methodology for choosing the best quantum size. Our initial measure was to choose a quantum size that is roughly 0.015%–0.1% of the total expected deviation (the absolute value of the range of the quantity in the reference simulation) of the quantized state variable. One objective of the work reported in this paper was to investigate the relationship between quantum size and error. Understanding this relationship could lead to a more rigorous methodology for choosing quantum size based on a desired error bound.

Figures 4–6 show the accuracy with which the LIQSS method tracks the reference method. The figures also show that atoms update asynchronously; over any particular period of time, each individual atom experiences a

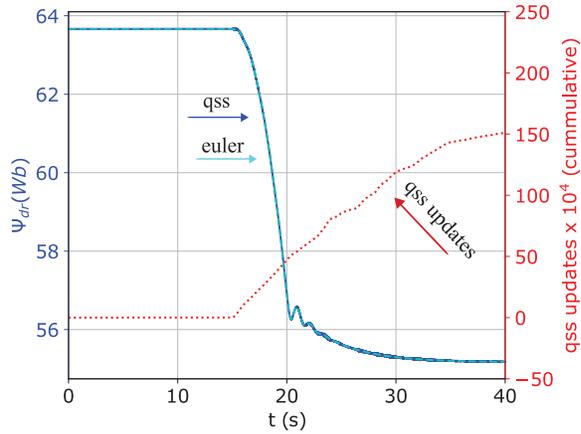

**Figure 4.** Rotor d-axis flux. The flux computed by the QSS method is nearly identical to that computed by the reference method so the two lines are nearly indistinguishable. Cumulative count of $\Psi_{dr}$ atom updates shows little activity prior to torque ramp, higher activity during torque ramp, and a return to little activity as new steady state is attained.

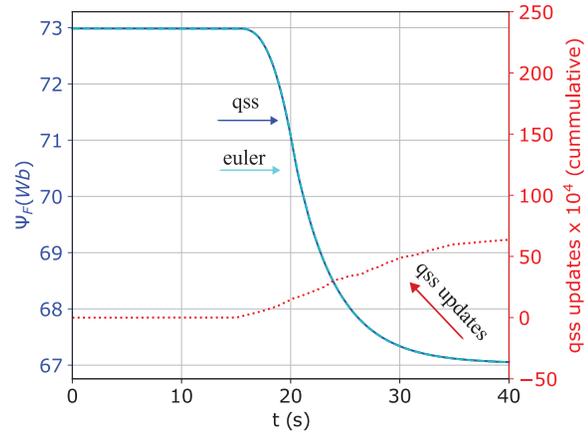

**Figure 6.** Field flux, showing good agreement between both computing methods and a total number of QSS atom updates that is smaller than the counts for d- and q-axis fluxes.

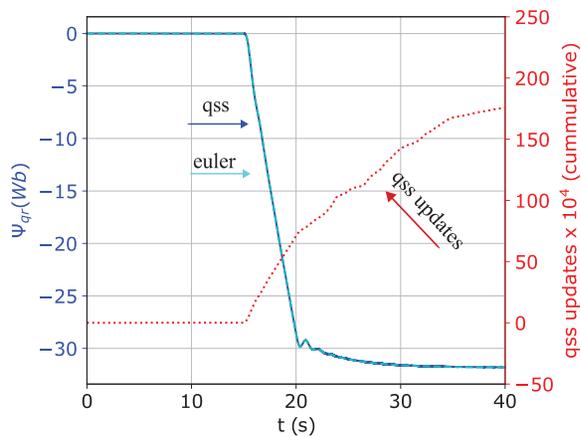

**Figure 5.** Rotor q-axis flux. The values computed by the QSS method and the reference method are nearly indistinguishable.

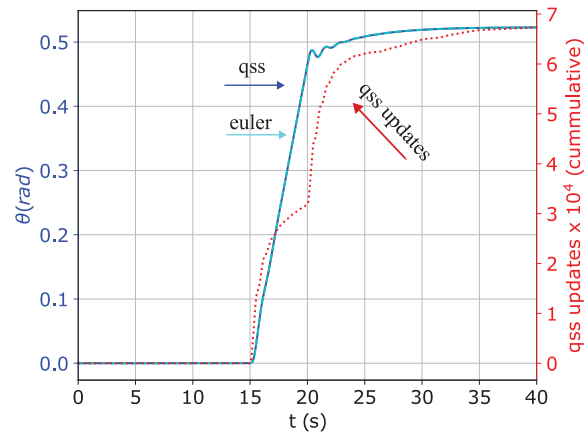

**Figure 7.** Rotor angle. QSS update rate shows interesting behavior with faster rates associated with beginning and ending of the torque ramp.

different number of updates. The cumulative number of updates for a particular atom is shown by the red lines in the following charts, where one can observe that any particular line reaches to a different number at the end of the simulation period.

An advantage of the QSS method over the reference method can be noted as the system reaches the new steady state condition and the rate of atom updates markedly decreases. For example, in Figure 4, during the interval from 0 to 15 s, while the system is in steady state, there are very few state updates. Then, during the period from 15 to 35 s, while the machine is accelerating and other states of the system are changing, the slope of the red line—representing the cumulative number of updates—is large. Then, finally after about 35 s, as the system approaches a new steady state, the update rate is again small. This aspect of the QSS method allows the simulation model to advance rapidly during steady state conditions. Furthermore, the cumulative number of updates depends on the chosen quantum size: A smaller quantum causes the system to require more frequent updates and hence the simulation advances slowly through time, while a larger quantum—up to a point—requires fewer updates and hence the simulation can advance in larger time increments. The effect of quantum size on simulation speed will be explored more fully in the next section.

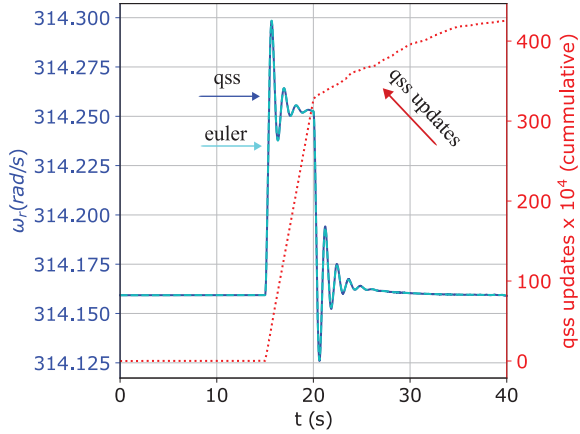

**Figure 8.** Rotor speed trajectory and update rates.

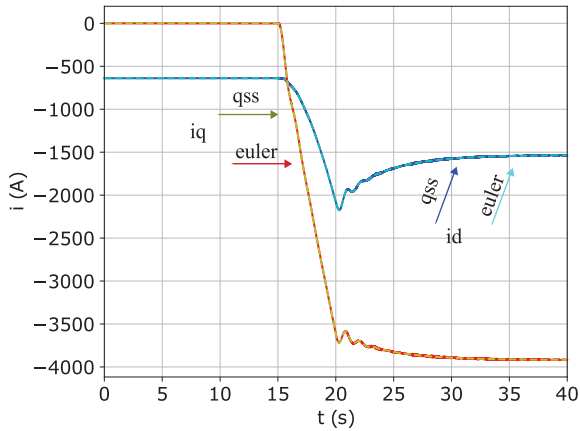

**Figure 9.** Comparisons of d- and q-axis rotor currents computed by the QSS and reference methods.

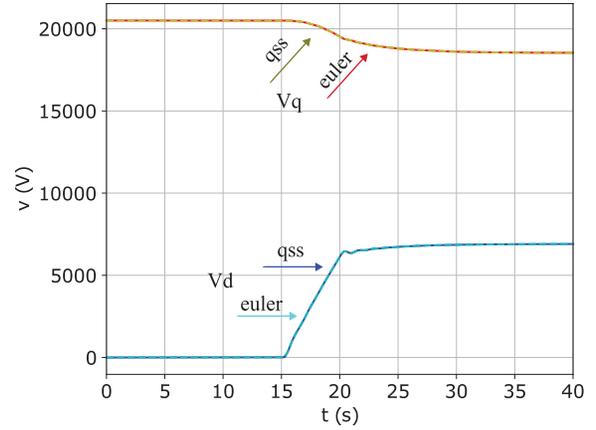

**Figure 10.** Comparisons of d- and q-axis voltages computed by the QSS and reference methods.

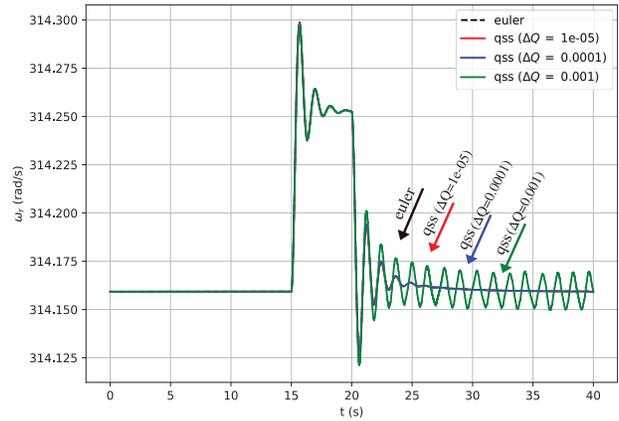

**Figure 11.** Rotor speed using different quantization sizes $\Delta Q = 10^{-5}$, $\Delta Q = 10^{-4}$, $\Delta Q = 10^{-3}$ versus Euler method reference solution.

The trajectory of rotor angle ($\theta$), as shown in Figure 7, is particularly interesting in that it shows how the count of this atom's updates increases immediately after the start of the torque ramp, then tapers off while the torque slew rate is constant during the interval between 15 and 20 s, then increases again at 20 s when the torque stops slewing and finally becomes small again as the rotor angle reaches its final steady state value. Figure 8 shows the trajectory of the rotor speed which always remains very close to $100 \times \pi$ rad/s but shows structure near the beginning and ending of the torque ramp and a small speed increase during the torque ramp as required to advance the rotor angle. In a later section of this paper, we will describe how both the accuracy and the error of rotor speed vary with choice of quantum size. The variables plotted in Figures 9 and 10 are functions of quantized states and therefore they do not have unique update rates.

## Accuracy and error analysis

Kofman[13] proved that for linear time invariant systems, the global error in QSS method can be bounded by a constant proportional to the quantum size. However, our reference system is highly non-linear so it is interesting to explore the behavior of the error with quantum size.

The effect of quantization size on simulation accuracy is shown in Figures 11–13, where several system variables are plotted for several different quantum sizes, with enhanced detail during particular time periods. A larger quantum size results in both a lower update rate and higher error amplitude compared to a situation with a smaller quantum size. In each case, the quoted quantum size applies to all of the state variables except rotor speed, for which the quantum is 1/10 that of the other variables. In each plot, the trajectory with green corresponds to the largest quantum size ($\Delta Q = 10^{-3}$), while blue and red correspond to smaller sizes ($\Delta Q = 10^{-4} and \Delta Q = 10^{-5}$, respectively). The rotor speed calculated with the largest quantum size has the largest error in comparison to the reference solution. This is evident in both Figure 12—in which resolution is increased near the apex of the speed trajectory—and Figure 13 where higher resolution shows a slightly oscillating speed trajectory after the torque ramp. These high frequency oscillations are inherent to the QSS method, and the amplitude of these oscillations increases as the quantum size increases.

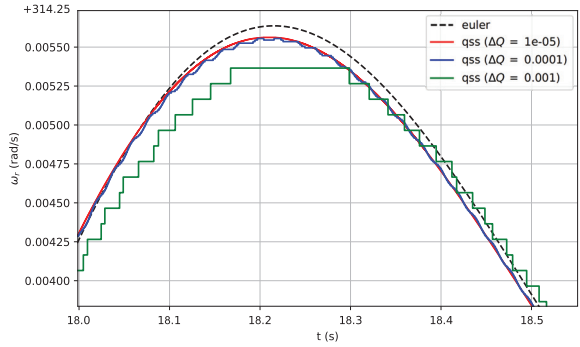

**Figure 12.** Zoom plots from 18 to 18.5 s of rotor speed using different quantization sizes $\Delta Q = 10^{-5}$, $\Delta Q = 10^{-4}$, $\Delta Q = 10^{-3}$ versus Euler method reference solution.

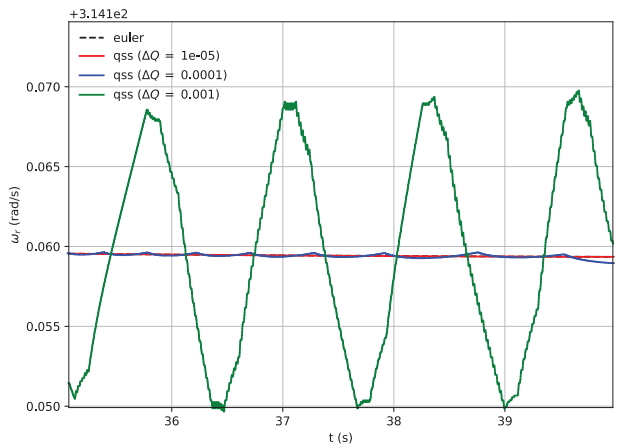

**Figure 13.** Zoom-in plots that show details for steady state situation of rotor speed using different quantization sizes $\Delta Q = 10^{-5}$, $\Delta Q = 10^{-4}$, $\Delta Q = 10^{-3}$ versus Euler reference solution. The oscillations have very small amplitude.

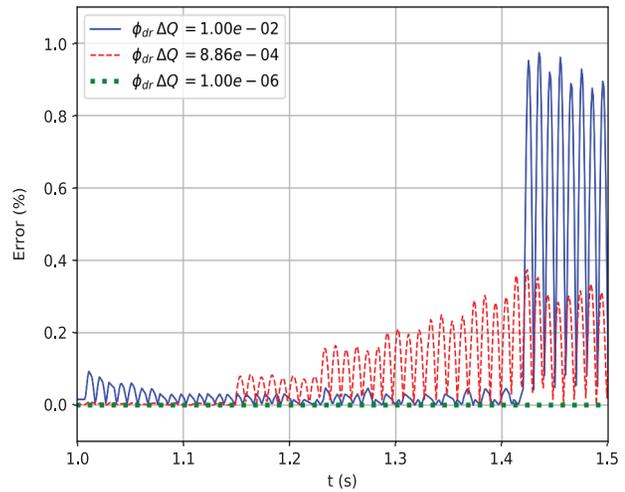

**Figure 14.** Error between reference solution and QSS solution of rotor d-axis flux for several different quantization sizes $\Delta Q = 10^{-2}$, $\Delta Q = 8.86 \times 10^{-4}$, and $\Delta Q = 10^{-6}$.

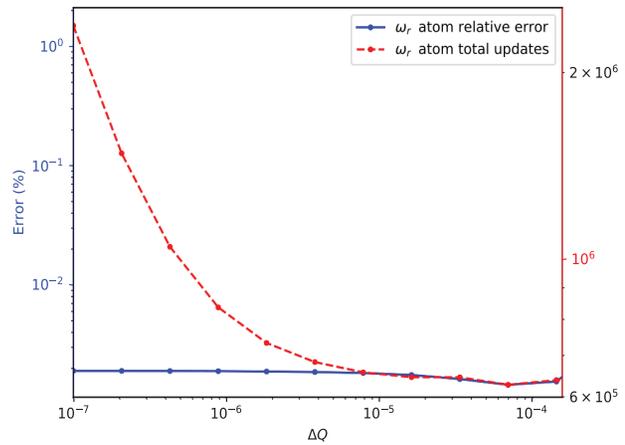

**Figure 15.** Rotor speed updates versus relative error when rotor speed quantum is set to $\Delta Q = 10^{-7}$ and the quantization size of the rest of the system variables is $\Delta Q = 10^{-4}$. Possibly unnecessary high precision with low benefit of error reduction.

Although there appears to be a predictable relationship between error and quantum size over a certain range of quantum sizes, it is not known how to specify the appropriate quantum size for a desired error for any particular QSS model description. The sensitivity data provided in this paper have been empirically determined from simulation of this specific system with specific component parameters. Although this empirical data do provide useful insight into the relationship between quantum size and error, it does not solve the problem for the general case.

Figures 14 and 15 describe the same behavior as was shown in Figures 11–13 but from the error perspective.

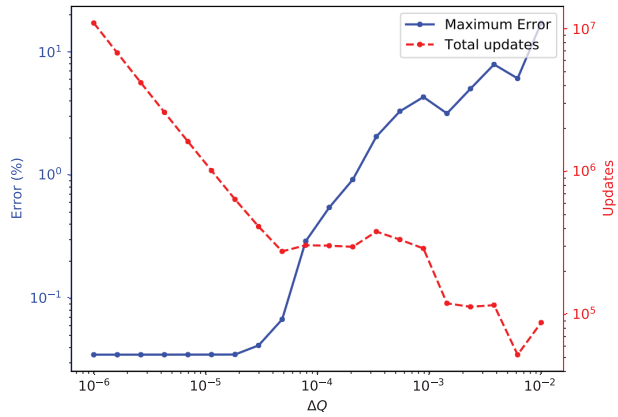

**Figure 16.** Maximum error of all atoms for different quantum sizes. Total number of the updates decreases as the system is simulated with bigger quantum sizes at the expense of increasing the error.

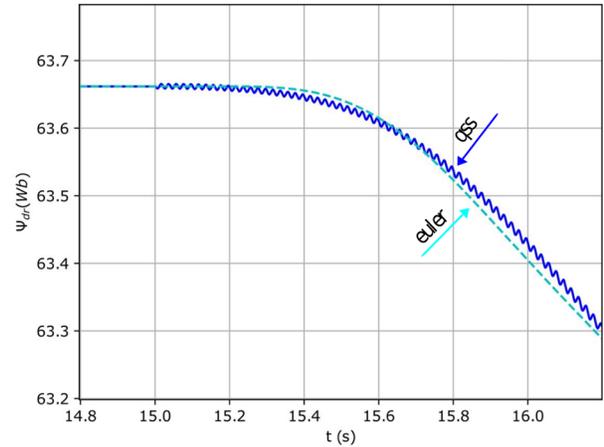

**Figure 17.** High-resolution plot showing the ripples in QSS solution of the rotor d-axis flux $\Psi_{dr}$ with quantum size of $\Delta Q = 10^{-4}$.

Figure 14 shows how the pointwise absolute error—the difference between the QSS simulation and the reference simulation—varies over a particular half-second interval. The time variation of error in the state variable $\Psi_{dr}$ is shown for several different quantum sizes ranging from $\Delta Q = 10^{-6}$ to $\Delta Q = 10^{-2}$ per unit. Clearly, a bigger quantum size produces a bigger error, but the relationship was not linear. Before the torque ramps up, the three different quantizations all produce negligible errors. After the torque starts to ramp up, the number of updates starts to grow and the models using larger quantization sizes produce larger errors. Although, a small quantization size does improve the simulation accuracy (smaller error amplitude), it also causes a larger model update rate. So, if computing speed is important, and if say, 1% error is tolerable, one might choose a large quantum size like $\Delta Q = 10^{-2}$ to achieve the requisite computing speed.

To generate the data shown in Figure 15, we quantized the rotor speed at $\Delta Q = 10^{-7}$ and the other state variables at $\Delta Q = 10^{-4}$. This was an experiment to see if choosing a relatively small quantization size for some particular interesting state, but leaving other quanta larger, would produce fewer updates for the whole system and still a small error for the particular state of interest. Here, the error is calculated according to Equation (1). The error is not invariant to the chosen system parameters. It is unknown how a different simulation scenario (i.e. a different set of model parameters) could affect the error. Figure 15 shows that the state of rotor speed experienced a high number of updates but the error did not reduce compared to choosing $\Delta Q = 10^{-4}$ for the whole system which produced the same output with a smaller number of updates. So, we suggest to keep the whole system at a unified quantum size when state variables have comparable magnitudes—except those that are derivatives of other states—instead of choosing any particular quantum size very small and the rest of the quantum sizes relatively larger.

Figure 16 shows the maximum error among any system variable at any time during the simulation interval. The graph is plotted on a logarithmic scale as a function of the quantum size, which was varied from $\Delta Q = 10^{-6}$ to $\Delta Q = 10^{-2}$ Wb. Also plotted is the corresponding sum of all updates of all atoms over the entire simulation. For small quantum sizes of $\Delta Q = 10^{-6}$ to $\Delta Q = 10^{-5}$ Wb, the error is very small and independent of quantum size. A logarithmic scale is used to emphasize that for very small quantum sizes ($\Delta Q < 10^{-5}$ Wb), a decrease in quantum size does not improve the accuracy, but it does impose a penalty on computational intensity (simulation update rate); the simulation takes longer to advance through time with no benefit in error reduction. A sweet spot is evident at quantum size between $\Delta Q = 10^{-5}$ and $\Delta Q = 10^{-4}$, where computational intensity has become relatively low while error also remains low. Above quantum size of $10^{-4}$, the error increases rapidly, but without concomitant reductions in computational intensity.

Although the QSS method does accurately track the reference solution, the method does inherently exhibit single-quantum oscillations. These oscillations are evident in Figure 17, which shows the direct axis rotor flux at high resolution just near the onset of the torque ramp at 15 s. The amplitudes of these high frequency oscillations reduce as the quantum size is reduced, but the oscillations are always present. Each oscillation represents an update event, so reducing the oscillation size comes at the expense of computation time. Figures 17 and 18 show these high frequency oscillations at quantum sizes of $\Delta Q = 10^{-4}$ and $\Delta Q = 10^{-5}$, respectively. This is an inherent nature of the QSS method and cannot be avoided.[5]

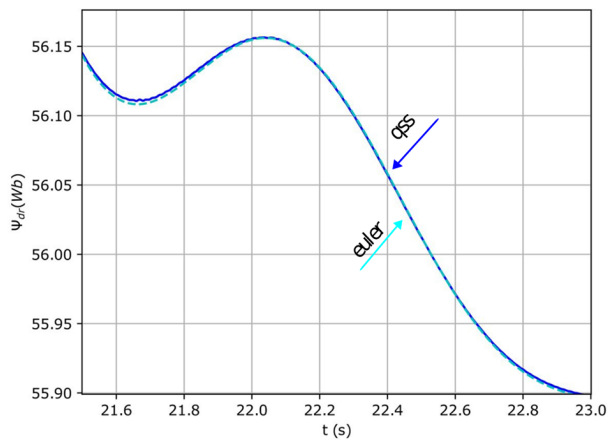

**Figure 18.** High-resolution plot showing the ripples in the QSS solution of the rotor d-axis flux $\Psi_{dr}$ with quantum size of $\Delta Q = 10^{-5}$. The amplitude of the ripples is negligible here compared to $\Delta Q = 10^{-4}$.

## Conclusion

The performance of the LIQSS1 method for analyzing the dynamics of power networks has been characterized by the simulation of a reference system. Uniform quantization of system state variables at 0.01% was found to yield accuracy within 0.4% of that achieved with a conventional state-space solution, but with a significant advantage in computational intensity, especially for systems that operate for long time in a quasi-steady state. Since the LIQSS method enables the user to individually set the quantum sizes of each state, we evaluated the performance as a function of quantization size. When the system was simulated using one uniform quantization size for all states, the total number of state updates decreased as quantum size increased, but above a quantization size of about $10^{-4}$, further increase in the quantization size did not significantly reduce the computational cost but it did decrease the simulation accuracy. When the quantization size for a single state of the interest was set smaller than the uniform quantization size of other states, refining the quantization size of that particular state variable did not necessarily improve the error of that particular state, and it did logarithmically increase the state update intensity of the system. Our observations of the effects of quantization size are limited by the particular system that was studied; other systems, especially those having state variables that are widely different in magnitude, may behave differently.

## Funding

The author(s) disclosed receipt of the following financial support for the research, authorship, and/or publication of this article: This work was supported in part by the US Office of Naval Research under grant N00014-16-1-2956.

## ORCID iD

Roger A Dougal  https://orcid.org/0000-0001-6152-1799

## Author biographies

**Roger A Dougal** is a Carolina Distinguished Professor of Electrical Engineering at the University of South Carolina. His research interests include power electronics, power and energy systems, and microgrids, with special interest in applications in Naval systems, and simulation and analysis tools that support design and analysis of those systems.

**Navid Gholizadeh** is a PhD student at the University of South Carolina. His research interests include Quantized Discrete Event System Specification (QDEVS) and Simulation of Power Systems. He earned his Masters in Electrical Engineering from Politecnico di Milano, Italy where his main focus was on IEC 61850 protocol for standardizing communications between monitoring devices. His Bachelors degree in Electrical Engineering was from Azad University-South, Tehran Branch, in Iran.

**Joseph M Hood** is a PhD candidate at the University of South Carolina. He has worked as an engineer for Siemens, developing simulation methods and models for the PSSE power system simulator. His research interests include simulation methods for power systems, advanced electric machine modeling, and Discrete Event Specification (DEVS) methods applied to continuous systems. His email address is hood@cec.sc.edu